\documentclass[12pt,a4paper]{article}
\usepackage{amssymb,amsmath}
\usepackage{amsfonts}
\usepackage{a4}
\usepackage{enumerate,verbatim}

\newtheorem{theo}{Theorem}
\newtheorem{lem}{Lemma}
\newtheorem{cor}{Corollary}
\newtheorem{rk}{Remark}
\newtheorem{eg}{Example}
\begin{document}
 \baselineskip18pt	
	\title{\bf New Quantum codes from constacyclic codes over a general non-chain ring }
	\author{Swati Bhardwaj$^1$ , Mokshi Goyal$^2$\footnote{\tiny Corresponding author,  Email: mokshigoyal@pec.edu.in }~ and Madhu Raka$^1$ \vspace{2mm}\\
		 $^1$\footnotesize{\em Department of Mathematics, Panjab University, Chandigarh, INDIA}\\
$^2$\footnotesize{\em Department of Applied Sciences, Punjab Engineering College, Chandigarh, INDIA}\\
		\date{}}
	
\maketitle
\vspace{-10mm}
{\abstract{
\noindent		Let $q$ be a prime power and let $\mathcal{R}=\mathbb{F}_{q}[u_1,u_2, \cdots, u_k]/\langle f_i(u_i),u_iu_j-u_ju_i\rangle$ be a finite non-chain ring, where $f_i(u_i), 1\leq i \leq k$  are polynomials, not all linear, which split into distinct linear factors over $\mathbb{F}_{q}$. We characterize  constacyclic codes over the ring $\mathcal{R}$ and study quantum codes from these. As an application, some new and better quantum codes, as compared to the best known codes, are obtained.  We also prove that the choice of the polynomials $f_i(u_i),$ $1 \leq i \leq k$ is irrelevant while constructing quantum codes from constacyclic codes over $\mathcal{R}$, it depends only on their degrees. It is shown that there always exists Quantum MDS code $[[n,n-2,2]]_q$ for any $n$  with $\gcd (n,q)\neq 1.$\vspace{2mm} 
		
	}
	\section{ Introduction}
 	 Quantum error-correcting codes are used by quantum computers to ensure that quantum information is protected from communication noise. Studies have been conducted extensively on quantum error-correcting codes. The main focus is on developing better quantum codes based on recent advances in the field. In $1995$, Shor $\cite{shor}$ came up with scheme for reducing decoherence in quantum memory.  In $1996$, Steane $\cite{stean}$ proposed important structural properties of quantum error-correcting codes. Further in $1998$, Calderbank et al. $\cite{Calderbank}$ constructed quantum error-correcting codes from classical codes, called CSS construction. Consequently, quantum error-correcting codes started to develop rapidly.\vspace{2mm}
 	
 \noindent	The class of cyclic, constacyclic and skew-constacyclic codes over finite non-chain rings have  been useful in finding good quantum error-correcting codes. In $2013,$ Qian et al. $\cite{Qian}$ studied quantum codes from cyclic codes over $\mathbb{F}_2+u\mathbb{F}_2.$ Generalizing their work,  Ashraf et al. $\cite{ashraf2015}$ constructed quantum codes from codes over $\mathbb{F}_{p}+v\mathbb{F}_{p}$ and Dertli et al. $\cite{DCE}$ considered binary quantum codes from cyclic codes over the ring $\mathbb{F}_2 + u\mathbb{F}_2 + v\mathbb{F}_2 + uv\mathbb{F}_2.$ Further, Ashraf and Mohammad $\cite{AM}$ generalized their work over the ring $\mathbb{F}_q + u\mathbb{F}_q + v\mathbb{F}_q + uv\mathbb{F}_q$ to determine new non-binary quantum codes. Codes over several non-chain rings have been characterized by various authors. Goyal and Raka \cite{CCDS} studied cyclic codes over the non-chain ring $\mathbb{F}_p[u]/\langle u^m-u\rangle$ and negacyclic codes \cite{DiscGR} over the non-chain ring $\mathbb{F}_q[u]/\langle f(u)\rangle$, where $f(u)$  splits into distinct linear factors over $\mathbb{F}_{q}$. Further  the cyclic and  constacyclic codes over the non-chain ring $\mathbb{F}_{q}[u,v]/$ $\langle f(u),g(v),uv-vu\rangle$, where $f(u)$ and $g(v)$ are non-linear polynomials and split into distinct linear factors over $\mathbb{F}_{q}$ have been discussed in detail by Goyal and Raka \cite{GR4}, \cite{GR5}. Bhardwaj and Raka \cite{BR} studied skew-constacyclic codes over it. On taking some special expression for polynomials $f(u)$ and $g(v)$, (e.g. $f(u)=u^2-u,  g(v)=v^2-1$) many researchers have obtained new  quantum error-correcting codes using CSS construction,  see $ \cite{ashraf2016}, \cite{constajaa}, \cite{constaadditive}, \cite{consta39}, \cite{MaGao}$. Recently, Islam et al. \cite{consta29}, \cite{consta35} and \cite{verma} worked over the rings $\mathbb{F}_p[u,v,w]/\langle u^2-1,v^2-1,w^2-1,uv-vu,vw-wv,wu-uw \rangle$ and $\mathbb{F}_{p^m}[u_1,u_2, \cdots, u_k]/\langle u_i^2-1,u_iu_j-u_ju_i \rangle$.  Note that $u_i^2-1=(u_i-1)(u_i+1)$ splits over $\mathbb{F}_q$. \vspace{2mm}

  \noindent Generalizing all these non-chain rings, here we discuss constacyclic codes over the ring $\mathcal{R}=$ $\mathbb{F}_{q}[u_1,u_2, \cdots, u_k]/$ $\langle f_i(u_i),u_iu_j-u_ju_i\rangle_{1\leq i,j\leq k}$, where $f_i(u_i), 1\leq i \leq k$  are polynomials of degree $m_i$, which split into distinct linear factors over $\mathbb{F}_{q}$. As an application, some new MDS or Almost MDS quantum codes such as  $[[18,14,3]]_{19}$, $[[24,22,2]]_{289}$, $[[22,14,4]]_{11}$,  $[[26,20,3]]_{13}$, $[[12,6,3]]_{19}$, $[[10,4,3]]_{25}$ and $[[24,18,3]]_{29}$ are obtained. Some better than previously known quantum codes such as $[[80,68,3]]_5$, $[[126,114,3]]_{7}$, $[[30,16,5]]_{11}$, $[[66,56,4]]_{11}$, $[[56,48,3]]_{49}$ and $[[88,80,3]]_{121}$
  are also obtained. See Tables 1, 2 and 3. While constructing quantum codes from constacyclic codes over $\mathcal{R}$, we observe that the choice of the polynomials $f_i(u_i), 1\leq i\leq k$ is irrelevant, it depends only on their degrees $m_i$ (Section \ref{sec5}). It is also shown that there always exists Quantum MDS code $[[n,n-2,2]]_q$ for any $n$  with $\gcd (n,q)\neq 1$ (Theorem \ref{th7}). \vspace{2mm}
 	
 	\noindent The paper is organized as follows: In Section 2, we discuss the ring $\mathcal{R},$ constacyclic codes over it and define a Gray map from the ring $\mathcal{R}$ to $\mathbb{F}_q^{m_1m_2\cdots m_k}$. In Section 3, we study quantum codes over $\mathbb{F}_q$ from constacyclic codes over  $\mathcal{R}$
 	  and obtain  some   new and better quantum codes  by using MAGMA computation system. The irrelevance of the expression of polynomials $f_i(u_i)$ for the construction of Quantum error-correcting codes is discussed in Section 4.  In Section 5, we conclude the paper.

\section{\bf Constacyclic codes over the ring $\mathcal{R}$ and the Gray map}

Let $R$ be a commutative ring with identity. An $R$-linear code $\mathcal{C}$ of length $n$ is an $R$-submodule of $R^n$. $\mathcal{C}$ is called $\lambda$-constacyclic code for a unit $\lambda$ in $R$ if it is invariant under the $\lambda$-constacyclic shift $\sigma_\lambda$, where\vspace{2mm} \\ $~~~~~~~~~~~~~~~~~~~~~~~~~~~\sigma_\lambda\big((c_0,c_1,...,c_{n-1})\big)=(\lambda c_{n-1},c_0,c_1,...,c_{n-2}).$\vspace{2mm}\\
If $\lambda=1$, $\mathcal{C}$ is cyclic. If $\lambda=-1$, $\mathcal{C}$ is called negacyclic. A $\lambda$-constacyclic code $\mathcal{C}$ of length $n$ over $R$ can be regarded as an ideal of 	 $R[x]/\langle x^n-\lambda\rangle$ under the correspondence \vspace{2mm}

$~~~~~~~~~~~~~~c=(c_0,c_1,...,c_{n-1})\to c(x)=c_0+c_1x+...+c_{n-1}x^{n-1}({\rm mod ~}x^n-\lambda)$. \vspace{2mm}

\noindent For a linear code $\mathcal{C}$  over $R$, the dual code  $\mathcal{C}^\bot$ is defined as $\mathcal{C}^\bot =\{x\in R^n~ |~ x \cdot y=0 ~ {\rm for ~ all~} y \in \mathcal{C}\}$, where $x\cdot y$ denotes the usual Euclidean inner product. If $\mathcal{C}$ is $\lambda$-constacyclic code over $R$, then $\mathcal{C}^\bot$ is a $\lambda^{-1}$-constacyclic code over $R$.\vspace{2mm}

\noindent If $R=\mathbb{F}_q$, a constacyclic code $\mathcal{C}$ of length $n$  has a unique monic generator polynomial $g(x)$ satisfying $g(x)|( x^n-\lambda)$. Let $x^n-\lambda=g(x)h(x)$ and  $h(x)=\sum_{i=0}^{n-r}h_ix^i$. Define the reciprocal of $h(x)$ as $h^\perp(x)=h_0^{-1}\big( h_{n-r}+h_{n-r-1}x+\cdots +h_0x^{n-r}\big)$. Then the dual code $\mathcal{C}^\bot$ is generated by $ h^\perp(x)$.
	
\subsection{The ring $\mathbb{F}_{q}[u_1,u_2, \cdots, u_k]/$ $\langle f_i(u_i),u_iu_j-u_ju_i\rangle$}
Let $q$ be a prime power, $q = p^{e}$, $\mathcal{R}=\mathbb{F}_{q}[u_1,u_2, \cdots, u_k]/$ $\langle f_i(u_i),u_iu_j-u_ju_i\rangle$ be a commutative ring, where $f_i(u_i), 1\leq i \leq k$  are polynomials of degree $m_i$, which split into distinct linear factors over $\mathbb{F}_q$.
Let
\begin{equation}\begin{array}{l}f_1(u_1)=(u_1-{\alpha}_{11})(u_1-{\alpha}_{12})\cdots(u_1-{\alpha}_{1m_1}),\\
f_2(u_2)=(u_2-{\alpha}_{21})(u_2-{\alpha}_{22})\cdots(u_2-{\alpha}_{2m_2}),\\ \cdots ~~~~~~~\cdots\\
f_i(u_i)=(u_i-{\alpha}_{i1})(u_i-{\alpha}_{i2})\cdots(u_i-{\alpha}_{im_i}),\\\cdots ~~~~~~~\cdots\\
f_k(u_k)=(u_k-{\alpha}_{k1})(u_k-{\alpha}_{k2})\cdots(u_k-{\alpha}_{km_k}),\end{array}\end{equation}

\noindent where $\alpha_{is_i} \in \mathbb{F}_q$, for $1\le s_i \le m_i, 1\leq i \leq k $ and $\alpha_{is_i} \neq \alpha_{is'_i}$, for all $s_i \neq s'_i$. $\mathcal{R}$ is a  non-chain ring of size ${q}^{m_1m_2\cdots m_k}$ and characteristic $p$. \\
For each $i$, $1 \leq i \leq k$, let $\epsilon_{s_i}^{(i)}$, $1\leq s_i\leq m_i$,  be elements of the ring $\mathcal{R}$ given by
\begin{equation} \begin{array}{ll}
\epsilon_{1}^{(i)}=\epsilon_{1}^{(i)}(u_i)= \frac{(u_i-\alpha_{i2})(u_i-\alpha_{i3})\cdots(u_i-\alpha_{is_i})\cdots(u_i-\alpha_{im_i})}
{(\alpha_{i1}-\alpha_{i2})(\alpha_{i1}-\alpha_{i3})\cdots(\alpha_{i1}-\alpha_{is_i})
	\cdots(\alpha_{i1}-\alpha_{im_i})}\\

\epsilon_{2}^{(i)}=\epsilon_{2}^{(i)}(u_i)= \frac{(u_i-\alpha_{i1})(u_i-\alpha_{i3})\cdots(u_i-\alpha_{is_i})\cdots(u_i-\alpha_{im_i})}
{(\alpha_{i2}-\alpha_{i1})(\alpha_{i2}-\alpha_{i3})\cdots(\alpha_{i2}-\alpha_{is_i})
	\cdots(\alpha_{i2}-\alpha_{im_i})}\\
\cdots ~~~~~\cdots\\
\epsilon_{s_i}^{(i)}=\epsilon_{s_i}^{(i)}(u_i)= \frac{(u_i-\alpha_{i1})(u_i-\alpha_{i2})\cdots(u_i-\alpha_{i(s_i-1)})(u_i-\alpha_{i(s_i+1)})\cdots(u_i-\alpha_{im_i})}{(\alpha_{is_i}-\alpha_{i1})(\alpha_{is_i}-\alpha_{i2})\cdots(\alpha_{is_i}-\alpha_{i(s_i-1)})
	(\alpha_{is_i}-\alpha_{i(s_i+1)})\cdots(\alpha_{is_i}-\alpha_{im_i})}\\\cdots ~~~~~~~\cdots \\
\epsilon_{m_i}^{(i)}=\epsilon_{m_i}^{(i)}(u_i)= \frac{(u_i-\alpha_{i1})(u_i-\alpha_{i2})\cdots(u_i-\alpha_{i(m_i-1)})}{(\alpha_{im_i}-\alpha_{i1})(\alpha_{im_i}-
\alpha_{i2})\cdots(\alpha_{im_i}-\alpha_{i(m_i-1)})}.
\end{array}\end{equation}
\noindent  If $m_i = 1$, for some $i$,  we define $\epsilon_{s_i}^{(i)}=1$.\vspace{2mm}

 \begin{lem}\label{lem1}\normalfont For each $i, 1\le i \le k$, we have  $\epsilon_{r_{i}}^{(i)}\epsilon_{s_{i}}^{(i)}\equiv 0({\rm mod ~} f_i(u_i)) {\rm ~for~} 1\leq r_{i},s_{i}\leq m_i,~r_{i} \neq s_{i}$, $(\epsilon_{s_i}^{(i)})^2\equiv\epsilon_{s_i}^{(i)}({\rm mod ~} f_i(u_i))$ and $\sum_{s_i=1}^{m_i} \epsilon_{s_i}^{(i)}\equiv 1 ({\rm mod ~} f_i(u_i))$.\end{lem}
		
\noindent {\bf Proof:} It is clear that $\epsilon_{r_{i}}^{(i)}\epsilon_{s_{i}}^{(i)}\equiv 0~ ({\rm mod ~} f_i(u_i)) {\rm ~for~} 1\leq r_{i},s_{i}\leq m_i,~r_{i} \neq s_{i}$.
To prove $(\epsilon_{s_i}^{(i)})^2\equiv\epsilon_{s_i}^{(i)} \pmod {f_i(u_i)}$, it is enough to prove that $\epsilon_{s_i}^{(i)}(\epsilon_{s_i}^{(i)}-1)\equiv 0 (\mod f_i(u_i))$.
So it is sufficient to prove that $(u_i-\alpha_{ir_i})|~\epsilon_{s_i}^{(i)}(\epsilon_{s_i}^{(i)}-1)$
for each $r_i, 1\le r_i \le m_i$. By definition of $\epsilon_{s_{i}}^{(i)}$, it is clear that $(u_i-\alpha_{ir_i})|~\epsilon_{s_{i}}^{(i)}$ for all $~r_i\neq s_i$.
 Also  $\epsilon_{s_{i}}^{(i)}(\alpha_{is_i})=1$,
 so  $(u_i-\alpha_{is_i})|~(\epsilon_{s_i}^{(i)}-1)$.
 Therefore   $f_i(u_i)|~\epsilon_{s_i}^{(i)}(\epsilon_{s_i}^{(i)}-1)$
 and hence $(\epsilon_{s_i}^{(i)})^2\equiv\epsilon_{s_i}^{(i)} \pmod {f_i(u_i)}$.
 As  $\epsilon_{1}^{(i)}(\alpha_{ir_i})+\epsilon_{2}^{(i)}(\alpha_{ir_i})+\cdots+\epsilon_{m_i}^{(i)}(\alpha_{ir_i})=1$ for all $r_i$, we find that  $(u_i-\alpha_{ir_i})|~\big(\epsilon_{1}^{(i)}(u_i)+\epsilon_{2}^{(i)}(u_i)+\cdots+
\epsilon_{m_i}^{(i)}(u_{i})-1\big)$ for all $r_i$ and hence $\epsilon_{1}^{(i)}+\epsilon_{2}^{(i)}+\cdots+
\epsilon_{m_i}^{(i)}\equiv 1 ({\rm mod ~} f_i(u_i))$.\hfill $\square$ \vspace{4mm}

\noindent	 For $1\le s_1\le m_1 $, $1\le  s_2\le m_2 $, $\cdots,1\le  s_k\le m_k $   define
$$ \eta_{s_1s_2\cdots s_k}= \epsilon_{s_1}^{(1)}(u_1)\epsilon_{s_2}^{(2)}(u_2)\cdots \epsilon_{s_k}^{(k)}(u_k) .$$
\noindent Working as in \cite{GR4}, we get

\begin{lem}\label{lem2}\normalfont For $ s_i= 1,2,\cdots,m_i, $ $1\leq i \leq k$, $\eta_{s_1s_2\cdots s_k}$'s are primitive central orthogonal idempotents of the ring $\mathcal{R}$.\end{lem}

\begin{eg}\normalfont Let $f_1(u_1)=u_1^2-1=(u_1-1)(u_1+1), f_2(u_2)=u_2^2-1, f_3(u_3)=u_3^2-1$. Then $\epsilon_1^{(1)}= \frac{1}{2}(u_1+1)$,  $\epsilon_2^{(1)}= \frac{1}{2}(1-u_1)$, $\epsilon_1^{(2)}= \frac{1}{2}(u_2+1)$, $\epsilon_2^{(2)}= \frac{1}{2}(1-u_2)$, $\epsilon_1^{(3)}= \frac{1}{2}(u_3+1)$,  $\epsilon_2^{(3)}= \frac{1}{2}(1-u_3)$. Therefore, \vspace{2mm}
\begin{equation*} \eta_{111}= \frac{1}{8}(1+u_1+u_2+u_3+u_1u_2+u_2u_3+u_3u_1+u_1u_2u_3) \end{equation*}
\begin{equation*} \eta_{211}= \frac{1}{8}(1-u_1+u_2+u_3-u_1u_2+u_2u_3-u_3u_1-u_1u_2u_3) \end{equation*}
\begin{equation*} \eta_{121}= \frac{1}{8}(1+u_1-u_2+u_3-u_1u_2-u_2u_3+u_3u_1-u_1u_2u_3) \end{equation*}
\begin{equation*} \eta_{112}= \frac{1}{8}(1+u_1+u_2-u_3+u_1u_2-u_2u_3-u_3u_1-u_1u_2u_3) \end{equation*}
\begin{equation*} \eta_{221}= \frac{1}{8}(1-u_1-u_2+u_3+u_1u_2-u_2u_3-u_3u_1+u_1u_2u_3) \end{equation*}
\begin{equation*} \eta_{212}= \frac{1}{8}(1-u_1+u_2-u_3-u_1u_2-u_2u_3+u_3u_1+u_1u_2u_3) \end{equation*}
\begin{equation*} \eta_{122}= \frac{1}{8}(1+u_1-u_2-u_3-u_1u_2+u_2u_3-u_3u_1+u_1u_2u_3) \end{equation*}
\begin{equation*} \eta_{222}= \frac{1}{8}(1-u_1-u_2-u_3+u_1u_2+u_2u_3+u_3u_1-u_1u_2u_3) \end{equation*}
 \noindent These are the same primitive idempotent as given in  \cite{consta29}.
\end{eg}

\noindent Throughout the paper by $\underset{s_1,s_2,\cdots,s_k}{\bigoplus}$ (or $\underset{s_1,s_2,\cdots,s_k}{\prod}$) we will mean that the direct sum (or the product) is over $k$ variables $s_1,s_2,\cdots,s_k$, where each $s_i$ varies from $1$ to $m_i$, $1 \le i \le k$. \vspace{2mm}

 \noindent The decomposition theorem of ring theory tells us that $$\mathcal{R}= \underset{s_1,s_2,\cdots,s_k}{\bigoplus}~\eta_{s_1s_2\cdots s_k}\mathcal{R}\cong \underset{s_1,s_2,\cdots,s_k}{\bigoplus}~\eta_{s_1s_2\cdots s_k}\mathbb{F}_q $$
\noindent Every element  $c$ in $\mathcal{R}$ can be uniquely expressed as $c=\underset{r_1,r_2,\cdots,r_k}{\bigoplus}~ \eta_{r_1r_2\cdots r_k}c_{r_1r_2\cdots r_k},$ \\where $c_{r_1r_2\cdots r_k} \in \mathbb{F}_{q}$. The same is true for elements of $\mathcal{R}^n$. Note that   $ c( \eta_{s_1s_2\cdots s_k})= \eta_{s_1s_2\cdots s_k}c_{s_1s_2\cdots s_k}$.\vspace{2mm}
		
\noindent For a linear code $\mathcal{C } $ of length $n$ over the ring $\mathcal{R}$, let for each pair $(s_1,s_2,\cdots,s_k), 1 \leq s_i \leq m_i, 1 \leq i \leq k$,  \vspace{2mm}

	$\begin{array}{ll}\mathcal{C }_{s_1s_2\cdots s_k}= & \Big\{ x_{s_1s_2\cdots s_k}\in \mathbb{F}_{q}^n : \exists ~x_{r_1r_2\cdots r_k}  \in \mathbb{F}_{q}^n, (r_1,r_2,\cdots, r_k)\neq (s_1,s_2,\cdots, s_k)\vspace{2mm}   \\  &{\rm ~such ~that~}\underset{r_1,r_2,\cdots, r_k}{\bigoplus}~\eta_{r_1r_2\cdots r_k}x_{r_1,r_2,\cdots, r_k} \in \mathcal{C}\Big\}.\end{array}$\vspace{2mm}\\
		Then $\mathcal{C }_{s_1s_2\cdots s_k}$ are linear codes of length $n$ over $\mathbb{F}_{q}$, and $$\mathcal{C}=\underset{s_1,s_2,\cdots,s_k}{\bigoplus}~\eta_{s_1s_2\cdots s_k}\mathcal{C}_{s_1s_2\cdots s_k},~~ {\rm i.e.~~} \eta_{s_1s_2\cdots s_k}\mathcal{C}= \eta_{s_1s_2\cdots s_k}\mathcal{C}_{s_1s_2\cdots s_k},$$ $$
		 |\mathcal{C }|= \underset{s_1,s_2,\cdots,s_k} {\prod}|\mathcal{C }_{s_1s_2\cdots s_k}|.$$
 	
\noindent Working as in \cite{BR}, we get

\begin{theo}\label{th1} Let $\mathcal{C}=\underset{s_1,s_2,\cdots,s_k}{\bigoplus}~\eta_{s_1s_2\cdots s_k}\mathcal{C}_{s_1s_2\cdots s_k}$ be a linear code of length $n$ over $\mathcal{R}$. Then
	
	\begin{enumerate}[$\rm(i)$]\item $\mathcal{C}^\perp=\underset{s_1,s_2,\cdots,s_k}{\bigoplus}~\eta_{s_1s_2\cdots s_k}\mathcal{C}^\perp_{s_1s_2\cdots s_k}$,
		
		\item $ \mathcal{C}^\perp \subseteq \mathcal{C}$  if and only if 	 $\mathcal{C}_{s_1s_2\cdots s_k}^\perp \subseteq\mathcal{C}_{s_1s_2\cdots s_k}$  and
		
		\item $|\mathcal{C}^\perp|= \underset{s_1,s_2,\cdots,s_k} {\prod}~|\mathcal{C}^\perp_{s_1s_2\cdots s_k}|$.
		
	\end{enumerate}\end{theo}

\subsection{Constacyclic codes over the ring $\mathcal{R}$}

Let $\lambda$ be an element in $\mathcal{R}$ given by \begin{equation}\label{eq3}\lambda = \underset{s_1,s_2,\cdots,s_k}{\bigoplus}~ \eta_{s_1s_2\cdots s_k}\lambda_{s_1s_2\cdots s_k}, ~~\lambda_{s_1s_2\cdots s_k} \in \mathbb{F}_{q}.\end{equation}
Clearly $\lambda$ is a unit in $\mathcal{R}$ if and only if $\lambda_{s_1s_2\cdots s_k}$ are units in $\mathbb{F}_{q}$, i.e., if and only if $\lambda_{s_1s_2\cdots s_k} \in \mathbb{F}_{q}\setminus\{0\}.$\vspace{2mm}

\noindent Note that $\lambda^2=1$ if and only if  $\lambda_{s_1s_2\cdots s_k}^2=1$, i.e., if and only if $\lambda_{s_1s_2\cdots s_k} = \pm 1$. \vspace{2mm}

\noindent The following Theorems 2, 3 and 4 are analogous to those of \cite{GR5}.

\begin{theo}  Let the unit $\lambda$ be as defined in {\normalfont(\ref{eq3})}. A linear code  $\mathcal{C}=\\\underset{s_1,s_2,\cdots,s_k}{\bigoplus}~\eta_{s_1s_2\cdots s_k}\mathcal{C}_{s_1s_2\cdots s_k}$ is a  $\lambda$-constacyclic code of length $n$  over $\mathcal{R}$ if and only if $\mathcal{C}_{s_1s_2\cdots s_k}$ are  $\lambda_{s_1s_2\cdots s_k}$-constacyclic codes of length $n$ over $\mathbb{F}_{q}$, for all $1 \leq s_i \leq m_i, 1\leq i \le k$.
  \end{theo}

  \noindent{\bf Proof:} Let $c=(c_0,c_1,\cdots,c_{n-1})\in \mathcal{C}$, where
  $c_r=\underset{s_1,s_2,\cdots,s_k}{\bigoplus}~\eta_{s_1s_2\cdots s_k}a_{s_1s_2\cdots s_k}^{(r)}$ for each $r,~ 0\le r \le n-1$. Let $a_{s_1s_2\cdots s_k}= ( a_{s_1s_2\cdots s_k}^{(0)},a_{s_1s_2\cdots s_k}^{(1)},\cdots,a_{s_1s_2\cdots s_k}^{(n-1)}),$ so that $c=\underset{s_1,s_2,\cdots,s_k}{\bigoplus}\eta_{s_1s_2\cdots s_k}a_{s_1s_2\cdots s_k}$, where $ a_{s_1s_2\cdots s_k}\in \mathcal{C}_{s_1s_2\cdots s_k}$. Using the properties of primitive idempotents $\eta_{s_1s_2\cdots s_k}$
  $$ \lambda c_{n-1}=  \Big(\underset{s_1,s_2,\cdots,s_k}{\bigoplus}~\eta_{s_1s_2\cdots s_k}\lambda_{s_1s_2\cdots s_k}\Big) \Big(\underset{s_1,s_2,\cdots,s_k}{\bigoplus}~\eta_{s_1s_2\cdots s_k}a_{s_1s_2\cdots s_k}^{(n-1)}\Big)$$
   $~~~~~~~~~~~~~~~~~=\underset{s_1,s_2,\cdots,s_k}{\bigoplus}~\eta_{s_1s_2\cdots s_k}\lambda_{s_1s_2\cdots s_k}a_{s_1s_2\cdots s_k}^{(n-1)}.$

  \noindent Therefore, $$\begin{array}{ll} \sigma_{\lambda}(c)&= (\lambda c_{n-1}, c_{0},\cdots,c_{n-2})\vspace{2mm}\\&= \Big(\underset{s_1,s_2,\cdots,s_k}{\bigoplus}~\eta_{s_1s_2\cdots s_k}\lambda_{s_1s_2\cdots s_k}a_{s_1s_2\cdots s_k}^{(n-1)}, \underset{s_1,s_2,\cdots,s_k}{\bigoplus}~\eta_{s_1s_2\cdots s_k}a_{s_1s_2\cdots s_k}^{(0)},\cdots,\\& ~~~~~~~~~\underset{s_1,s_2,\cdots,s_k}{\bigoplus}~\eta_{s_1s_2\cdots s_k}a_{s_1s_2\cdots s_k}^{(n-2)}\Big)\vspace{2mm}\\&= \underset{s_1,s_2,\cdots,s_k}{\bigoplus}~\eta_{s_1s_2\cdots s_k}\Big(\lambda_{s_1s_2\cdots s_k}a_{s_1s_2\cdots s_k}^{(n-1)}, a_{s_1s_2\cdots s_k}^{(0)},\cdots,a_{s_1s_2\cdots s_k}^{(n-2)}\Big)\vspace{2mm}\\
  &=\underset{s_1,s_2,\cdots,s_k}{\bigoplus}~\eta_{s_1s_2\cdots s_k}\sigma_{\lambda_{s_1s_2\cdots s_k}}(a_{s_1s_2\cdots s_k}).\end{array}$$

  \noindent Hence $\sigma_{\lambda}(c) \in \mathcal{C} $ if and only if $\sigma_{\lambda_{s_1s_2\cdots s_k}}(a_{s_1s_2\cdots s_k}) \in \mathcal{C}_{s_1s_2\cdots s_k}$.  \hfill $\square$\vspace{2mm}

  \begin{theo}  If   $\mathcal{C}=\underset{s_1,s_2,\cdots,s_k}{\bigoplus}~\eta_{s_1s_2\cdots s_k}\mathcal{C}_{s_1s_2\cdots s_k}$ is a $\lambda$-constacyclic code of length $n$  over $\mathcal{R}$, then $\mathcal{C}^\perp$ is  $\lambda^{-1}$-constacyclic code over $\mathcal{R}$ and $\mathcal{C}_{s_1s_2\cdots s_k}^\perp$ are $\lambda_{s_1s_2\cdots s_k}^{-1}$-constacyclic codes over $\mathbb{F}_{q}$, where $\lambda$ is as given in {\rm(\ref{eq3})}.  Further for $\mathcal{C}^\perp\subseteq \mathcal{C}$,  it is necessary that $\lambda^{-1}=\lambda$ i.e., $\lambda^2=1$ i.e., $ \lambda=  \sum_{s_1,s_2,\cdots,s_k} (\pm\eta_{s_1s_2\cdots s_k})$.\end{theo}

  \noindent{\bf Proof:} The first statement is a well-known result. Also by Theorem 1, we have $\mathcal{C}^\perp=\underset{s_1,s_2,\cdots,s_k}{\bigoplus}~\eta_{s_1s_2\cdots s_k}\mathcal{C}_{s_1s_2\cdots s_k}^\perp$, and $\lambda^{-1} = \underset{s_1,s_2,\cdots,s_k}{\bigoplus} \eta_{s_1s_2\cdots s_k}\lambda_{s_1s_2\cdots s_k}^{-1}$. Therefore $\mathcal{C}_{s_1s_2\cdots s_k}^\perp$ are $\lambda_{s_1s_2\cdots s_k}^{-1}$-constacyclic codes over $\mathbb{F}_{q}$. Further $\mathcal{C}^\perp\subseteq \mathcal{C}$ if and only if $\mathcal{C}^\perp_{s_1s_2\cdots s_k}\subseteq \mathcal{C}_{s_1s_2\cdots s_k}$. Now for $\mathcal{C}_{s_1s_2\cdots s_k}$ to be  dual containing, it is necessary that  $\lambda_{s_1s_2\cdots s_k}=\lambda_{s_1s_2\cdots s_k}^{-1}$ in $\mathbb{F}_{q}$, i.e.,  $\lambda_{s_1s_2\cdots s_k}=\pm 1$. \hfill $\square$\vspace{2mm}

\begin{theo}  Let  $\mathcal{C}=\underset{s_1,s_2,\cdots,s_k}{\bigoplus}~\eta_{s_1s_2\cdots s_k}\mathcal{C}_{s_1s_2\cdots s_k}$ be a  $\lambda$-constacyclic code  of length $n$ over $\mathcal{R}$.   Suppose that  $\lambda_{s_1s_2\cdots s_k}$-constacyclic codes  $\mathcal{C}_{s_1s_2\cdots s_k}$ are generated by $ g_{s_1s_2\cdots s_k}(x) $, where $x^n-\lambda_{s_1s_2\cdots s_k}=g_{s_1s_2\cdots s_k}(x) h_{s_1s_2\cdots s_k}(x)$ for $ ~1 \leq s_i \leq m_i, i \leq i \leq k$. Then there exist  polynomials $g(x)$ and $h^\perp(x)$ in $\mathcal{R}[x]$ such that
	
	\begin{enumerate}[$\rm(i)$] \item $\mathcal{C}=\langle g(x)\rangle,$
		
		\item  $g(x)$ is a  divisor of $(x^{n}-\lambda)$,
		
		\item  $|\mathcal{C }|=q^{m_1m_2...m_k n-\sum_{s_1,s_2,\cdots,s_k}deg(g_{s_1s_2\cdots s_k}(x))}$,
		\item $ \mathcal{C}^\perp=\langle h^\perp(x)\rangle,$ where
		$ h^\perp(x)=\sum_{s_1,s_2,\cdots,s_k}\eta_{s_1s_2\cdots s_k}h_{s_1s_2\cdots s_k}^\perp(x)$ and
		\item
		
		$ |\mathcal{C}^\perp|=q^{\sum_{s_1,s_2,\cdots,s_k}deg(g_{s_1s_2\cdots s_k}(x))}$.\end{enumerate}	
\end{theo}

\noindent{\bf Proof:} First we show that $\mathcal{C}$ is equal to $\mathcal{E}=\langle \eta_{s_1s_2\cdots s_k}g_{s_1s_2\cdots s_k}(x), 1\leq s_i\le m_i, 1\le i \le k\rangle$.
Let $c(x) \in \mathcal{C}$. Since $\mathcal{C}_{s_1s_2\cdots s_k}=\langle g_{s_1s_2\cdots s_k}(x) \rangle$ and $\mathcal{C}=\underset{s_1,s_2,\cdots, s_k}{\bigoplus}~\eta_{s_1s_2\cdots s_k}\mathcal{C}_{s_1s_2\cdots s_k}$, we have  $c(x)=\underset{s_1,s_2,\cdots, s_k}{\sum}~\eta_{s_1s_2\cdots s_k}u_{s_1s_2\cdots s_k}(x)g_{s_1s_2\cdots s_k}(x)$ for some $u_{s_1s_2\cdots s_k}(x) \in
\mathbb{F}_q[x]$. Therefore, $c(x)\in \mathcal{E}$ and so $\mathcal{C} \subseteq \mathcal{E}$. \vspace{2mm}

\noindent Conversely, let $c(x)=\underset{s_1,s_2,\cdots, s_k}{\sum}~\eta_{s_1s_2\cdots s_k}f_{s_1s_2\cdots s_k}(x)g_{s_1s_2\cdots s_k}(x) \in \mathcal{E}$, where\\ $f_{s_1s_2\cdots s_k}(x) \in
\mathcal{R}[x]$. As $\mathcal{R}= \underset{s_1,s_2,\cdots, s_k}{\bigoplus}~\eta_{s_1s_2\cdots s_k}\mathbb{F}_q$, each $f_{s_1s_2\cdots s_k}(x)= \\\underset{s_1,s_2,\cdots, s_k}{\sum}~\eta_{s_1s_2\cdots s_k}u_{s_1s_2\cdots s_k}(x)$ for some $u_{s_1s_2\cdots s_k}(x) \in
\mathbb{F}_q[x]$. Now $\eta_{s_1s_2\cdots s_k}f_{s_1s_2\cdots s_k}(x)\\=\eta_{s_1s_2\cdots s_k}u_{s_1s_2\cdots s_k}(x)$ as $\eta_{s_1s_2\cdots s_k}$ are primitive orthogonal idempotents. So we get that $c(x)=\underset{s_1,s_2,\cdots, s_k}{\sum}~\eta_{s_1s_2\cdots s_k}u_{s_1s_2\cdots s_k}(x)g_{s_1s_2\cdots s_k}(x)\in \underset{s_1,s_2,\cdots, s_k}{\bigoplus}~\eta_{s_1s_2\cdots s_k}\langle g_{s_1s_2\cdots s_k}(x)\rangle\\ = \mathcal{C}$, hence $ \mathcal{C} = \mathcal{E}$. \vspace{2mm}

\noindent Let $g(x)= \sum_{s_1,s_2,\cdots, s_k} \eta_{s_1s_2\cdots s_k}g_{s_1s_2\cdots s_k}(x)$. Then clearly $\langle g(x) \rangle \subseteq \mathcal{E}=\mathcal{C}$. On the other hand  $\eta_{s_1s_2\cdots s_k}g(x)=\eta_{s_1s_2\cdots s_k}g_{s_1s_2\cdots s_k}(x)$, so $\mathcal{C}\subseteq \langle g(x) \rangle$. \vspace{2mm}

\noindent Let $ h(x)= \underset{s_1,s_2,\cdots, s_k}{\sum}~\eta_{s_1s_2\cdots s_k}h_{s_1s_2\cdots s_k}(x),$ then one finds that
$h(x) g(x)=x^n-\lambda$, so $g(x)$  divides  $x^n-\lambda$.\vspace{2mm}

\noindent Since $|\mathcal{C }|= \underset{s_1,s_2,\cdots, s_k} {\prod}|\mathcal{C }_{s_1s_2\cdots s_k}|$ and $|\mathcal{C }_{s_1s_2\cdots s_k}|=q^{ n-deg(g_{s_1s_2\cdots s_k}(x))}$, we get (iii).\vspace{2mm}

\noindent Let $\mathcal{C}_{s_1s_2\cdots s_k}^\perp= \langle h_{s_1s_2\cdots s_k}^\perp(x) \rangle$ and take $h^\perp(x)= \sum_{s_1,s_2,\cdots, s_k} \eta_{s_1s_2\cdots s_k}h_{s_1s_2\cdots s_k}^\perp(x)$. Since $ \mathcal{C}^\perp=\underset{s_1,s_2,\cdots, s_k}{\bigoplus}~\eta_{s_1s_2\cdots s_k}\mathcal{C}_{s_1s_2\cdots s_k}^\perp,$  we get $\mathcal{C}^\perp= \langle h^\perp(x) \rangle$. (v) follows because $|\mathcal{C }||\mathcal{C^\perp }|= q^{m_1m_2\cdots m_kn}$. \hfill $\square$\vspace{2mm}

\begin{theo}\label{thm1}
	Let $\mathcal{C}=\underset{s_1,s_2,\cdots,s_k}{\bigoplus}~\eta_{s_1s_2\cdots s_k}\mathcal{C}_{s_1s_2\cdots s_k}$ be a $\lambda$-constacyclic code of length $n$ over $\mathcal{R}$, where $\mathcal{C}_{s_1s_2\cdots s_k}$ are $\lambda_{s_1s_2\cdots s_k}$-constacyclic codes over $\mathbb{F}_q$ generated by $g_{s_1s_2\cdots s_k} (x)$, where $\lambda_{s_1s_2\cdots s_k}=\pm 1$.  Then $\mathcal{C}^\perp \subseteq \mathcal{C}$ if and only if $\mathcal{C}_{s_1s_2\cdots s_k}^\perp \subseteq \mathcal{C}_{s_1s_2\cdots s_k}$, i.e., if and only if~ $$ x^n-\lambda_{s_1s_2\cdots s_k} \equiv 0 \pmod {g_{s_1s_2\cdots s_k} (x)g_{s_1s_2\cdots s_k}^\perp (x)}, $$  for $1\leq s_i\leq m_i, 1 \leq i \leq k.$
	
\end{theo}
\noindent \textbf{Proof:} We have  $ x^n-\lambda_{s_1s_2\cdots s_k}= g_{s_1s_2\cdots s_k}(x)h_{s_1s_2\cdots s_k}(x)$ over $\mathbb{F}_q$. By Theorem \ref{th1}, $\mathcal{C}^\perp \subseteq \mathcal{C}$ if and only if
$\mathcal{C}_{s_1s_2\cdots s_k}^\perp \subseteq \mathcal{C}_{s_1s_2\cdots s_k}$ which is so if and only if  for each $s_i$, $1\leq s_i\leq m_i, 1 \leq i \leq k$,
\begin{equation*} \begin{array}{ll}& g_{s_1s_2\cdots s_k}(x) \big|h_{s_1s_2\cdots s_k}^\perp(x)\vspace{2mm}\\\Leftrightarrow & g_{s_1s_2\cdots s_k}^\perp(x) \big|h_{s_1s_2\cdots s_k}(x)\vspace{2mm}\\\Leftrightarrow & g_{s_1s_2\cdots s_k}(x)g_{s_1s_2\cdots s_k}^\perp(x)\big|g_{s_1s_2\cdots s_k}(x)h_{s_1s_2\cdots s_k}(x)\vspace{2mm}\\\Leftrightarrow & g_{s_1s_2\cdots s_k}(x)g_{s_1s_2\cdots s_k}^\perp(x)\big|x^n-\lambda_{s_1s_2\cdots s_k}.\end{array}\end{equation*}

\begin{rk}\normalfont If $G_{s_1s_2\cdots s_k}$'s are generator matrices of codes $\mathcal{C}_{s_1s_2\cdots s_k}$ over $\mathbb{F}_q$, then a generator matrix of $\mathcal{C}$  is \vspace{2mm}

\[\begin{bmatrix}
	\eta_{11\cdots 1}G_{11\cdots 1} \\
	\eta_{21\cdots 1}G_{21\cdots 1}  \\
	\cdots \\
	\eta_{m_11\cdots 1}G_{m_11\cdots 1} \\
	\eta_{1 2 1\cdots 1}G_{12 1\cdots 1} \\
	\cdots \\
	\eta_{m_12 1\cdots 1}G_{m_12 1\cdots 1} \\
	\cdots \\
	\eta_{1m_2 1\cdots 1}G_{1m_2 1\cdots 1} \\\cdots\\
	\cdots\\\eta_{1m_2 \cdots m_k}G_{1m_2 \cdots m_k} \\\cdots\\
	\eta_{m_1m_2\cdots m_k}G_{m_1m_2\cdots m_k} \\
\end{bmatrix}.\]
 \end{rk}

\subsection{ The Gray map}

\noindent Every element $r(u_1,u_2,\cdots,u_k)$ of the ring $\mathcal{R}$ can be uniquely expressed as
$$ r(u_1,u_2,\cdots,u_k) = \underset{s_1,s_2,\cdots,s_k}{\bigoplus}~\eta_{s_1s_2\cdots s_k}a_{s_1s_2\cdots s_k},$$
where $a_{s_1s_2\cdots s_k} \in \mathbb{F}_q$ for $1 \leq s_i \leq m_i, 1 \leq i \leq k$. \\

\noindent Let $[a_{s_1s_2\cdots s_k}]_{s_1,s_2,\cdots,s_k}$ denote a $1\times m_1m_2\cdots m_k$ row matrix over $\mathbb{F}_q$ given by \\  $[a_{11\cdots 1},a_{21\cdots 1},\cdots a_{m_11\cdots 1}, a_{121\cdots 1},$  $\cdots, a_{m_121\cdots 1},\cdots,a_{1m_2 1\cdots 1},\cdots, a_{1m_2 \cdots m_k},\cdots, a_{m_1m_2\cdots m_k}]$. \vspace{2mm}

\noindent Define a Gray map $\Phi : \mathcal{R}\rightarrow \mathbb{F}_q^{m_1m_2\cdots m_k}$  by $$ r(u_1,u_2,\cdots,u_k)  \longmapsto[a_{s_1s_2\cdots s_k}]_{s_1,s_2,\cdots,s_k}V,$$
where $V$ is any non-singular matrix over $\mathbb{F}_q$ of order $m_1m_2\cdots m_k~\times~ m_1m_2\cdots m_k$.
This map can be extended from $\mathcal{R}^n$ to  $(\mathbb{F}_q^{m_1m_2\cdots m_k})^n$ component-wise. \\

	\noindent Let the Gray weight of an element $r \in \mathcal{R}$ be $w_{G}(r) =w_H(\Phi(r))$, the Hamming weight of $\Phi(r)$. The Gray weight of  a codeword
	$c=(c_0,c_1,\cdots,c_{n-1})$ $\in \mathcal{R}^n$ is defined as $w_{G}(c)=\sum_{i=0}^{n-1}w_{G}(c_i)=\sum_{i=0}^{n-1}w_H(\Phi(c_i))=w_H(\Phi(c))$. For any two elements $c_1, c_2 \in \mathcal{R}^n$, the Gray distance $d_{G}$ is given by $d_{G}(c_1,c_2)=w_{G}(c_1-c_2)=w_H(\Phi(c_1)-\Phi(c_2))$. \vspace{2mm}
	
	\noindent	The following theorem is analogous to a result of \cite{GR4}.

\begin{theo}\label{graydual} The Gray map $\Phi$ is an  $\mathbb{F}_q$-linear, one to one and onto map. It is also distance preserving map from ($\mathcal{R}^n$, Gray distance $d_{G}$) to $((\mathbb{F}_q^{m_1m_2\cdots m_k })^n$ Hamming distance $d_H$). Further, if the matrix $V$  satisfies $VV^T=\beta I$, $\beta \in \mathbb{F}_q^*$, where $V^T$ denotes the transpose of the matrix $V$, then $\Phi(\mathcal{C}^{\perp})=(\Phi(\mathcal{C}))^{\perp}$ for any linear code   $\mathcal{C}$  over $\mathcal{R}$. \end{theo}

\noindent \textbf{Proof.} The first two assertions hold as $V$ is an invertible matrix over $\mathbb{F}_q$.\\ Let now $V=[V_{1}^T, V_{2}^T, \cdots, V_{m_1m_2 \cdots m_k}^T] $, where\vspace{2mm}

\noindent  $V_t=[v^{(t)}_{11\cdots 1}, v^{(t)}_{21\cdots 1},\cdots, v^{(t)}_{m_11\cdots 1},v^{(t)}_{121\cdots 1},\cdots, v^{(t)}_{m_121\cdots 1},  \cdots,v^{(t)}_{1m_2 \cdots m_k},\cdots, v^{(t)}_{m_1m_2 \cdots m_k}]$\vspace{2mm} is a $1\times m_1m_2\cdots m_k$ matrix over $\mathbb{F}_q$, satisfying $VV^T=\beta I_{m_1m_2 \cdots m_k}$. So that
\begin{equation}\begin{array}{l} {\sum\limits_{t=1}^{m_1m_2 \cdots m_k}}~ (v_{s_1s_2 \cdots s_k}^{(t)})^2=\beta ~~{\rm for ~all ~} 1\leq s_i \leq m_i, \ 1\leq i \leq k {\rm ~~and ~~}\vspace{1mm}\\  {\sum\limits_{t=1}^{m_1m_2 \cdots m_k}}~ v_{s_1s_2 \cdots s_k}^{(t)}v_{s'_1s'_2 \cdots s'_k}^{(t)}=0 ~~{\rm for ~} (s_1s_2 \cdots s_k) \neq (s'_1s'_2 \cdots s'_k). \end{array}\end{equation}
Let $\mathcal{C}$ be a linear code over $\mathcal{R}$. Let $r=(r_0,r_1,\cdots,r_{n-1})\in \mathcal{C}^{\perp}$,  $u=(u_0,u_1,\cdots,u_{n-1})$
$\in \mathcal{C}$, where $r_i=\underset{s_1,s_2,\cdots,s_k}{\bigoplus}~\eta_{s_1s_2\cdots s_k}a^{(i)}_{s_1s_2\cdots s_k}$ and $u_i=\underset{s_1,s_2,\cdots,s_k}{\bigoplus}~\eta_{s_1s_2\cdots s_k}b^{(i)}_{s_1s_2\cdots s_k}$. So that $r\cdot u=0$. To prove that $\Phi(\mathcal{C}^{\perp})\subseteq(\Phi(\mathcal{C}))^{\perp}$, it is enough to prove that $\Phi(r)\cdot \Phi(u)=0 $. Using the properties of $\eta_{s_1s_2 \cdots s_k}$'s (Lemma \ref{lem2}), we get
$$ r_i u_i = \underset{s_1,s_2,\cdots,s_k}{\bigoplus}~\eta_{s_1s_2\cdots s_k}a^{(i)}_{s_1s_2\cdots s_k}b^{(i)}_{s_1s_2\cdots s_k}.$$ Then
$$\begin{array}{l}0=r\cdot u=\sum\limits_{i=0}^{n-1}r_iu_i= \sum\limits_{i=0}^{n-1}~\underset{s_1,s_2,\cdots,s_k}{\bigoplus}~\eta_{s_1s_2\cdots s_k}a^{(i)}_{s_1s_2\cdots s_k}b^{(i)}_{s_1s_2\cdots s_k}\\~~=\underset{s_1,s_2,\cdots,s_k}{\bigoplus}~\eta_{s_1s_2\cdots s_k}\Big( \sum\limits_{i=0}^{n-1}a_{s_1s_2\cdots s_k}^{(i)}\hspace{0.5mm}b_{s_1s_2\cdots s_k}^{(i)}\Big)\end{array} $$  implies that
\begin{equation}\begin{array}{l}{\sum\limits_{i=0}^{n-1}}a_{s_1s_2\cdots s_k}^{(i)}b_{s_1s_2\cdots s_k}^{(i)}=0, ~~~~~{\rm for ~ all ~} s_i, ~ 1 \leq s_i \leq m_i, 1 \leq i \leq k.\end{array}\end{equation}
Now
$$\begin{array}{l}\Phi(r_i)=[a^{(i)}_{s_1s_2\cdots s_k}]_{s_1,s_2,\cdots,s_k}V =[a^{(i)}_{s_1s_2\cdots s_k}]_{s_1,s_2,\cdots,s_k}[V_{1}^T, V_{2}^T, \cdots, V_{m_1m_2 \cdots m_k}^T]
\vspace{2mm}\\

=\Big[\underset{s_1,s_2,\cdots,s_k}{\bigoplus}a_{s_1s_2\cdots s_k}^{(i)}v_{s_1s_2\cdots s_k}^{(1)},\underset{s_1,s_2,\cdots,s_k}{\bigoplus}a_{s_1s_2\cdots s_k}^{(i)}v_{s_1s_2\cdots s_k}^{(2)},\cdots,\hspace{-1mm} \underset{s_1,s_2,\cdots,s_k}{\bigoplus}a_{s_1s_2\cdots s_k}^{(i)}v_{s_1s_2\cdots s_k}^{(m_1m_2 \cdots m_k)}\Big]\end{array}$$
Similarly  $\Phi(u_i)$ $$\begin{array}{l} =\Big[\underset{s'_1,s'_2,\cdots,s'_k}{\bigoplus}b_{s'_1s'_2 \cdots s'_k}^{(i)}v_{s'_1s'_2 \cdots s'_k}^{(1)},\underset{s'_1,s'_2,\cdots,s'_k}{\bigoplus}b_{s'_1s'_2 \cdots s'_k}^{(i)}v_{s'_1s'_2 \cdots s'_k}^{(2)},\cdots, \hspace{-1mm}\underset{s'_1,s'_2,\cdots,s'_k}{\bigoplus}b_{s'_1s'_2 \cdots s'_k}^{(i)}v_{s'_1s'_2 \cdots s'_k}^{(m_1m_2 \cdots m_k)}\Big]\end{array}$$
Using (4) and (5), we find that
$$\begin{array}{ll}\Phi(r)\cdot \Phi(u)\hspace{-0.5mm}&= {\sum\limits_{i=0}^{n-1}}\Phi(r_i)\cdot \Phi(u_i)\vspace{1mm}\\
&= {\sum\limits_{i=0}^{n-1}}\hspace{0.5mm}  {\sum\limits_{t=1}^{m_1m_2\cdots m_k}}\underset{s_1,s_2,\cdots,s_k}{\bigoplus}\underset{s'_1,s'_2,\cdots,s'_k}{\bigoplus}a_{s_1s_2\cdots s_k}^{(i)}\hspace{0.5mm}
b_{s'_1s'_2 \cdots s'_k}^{(i)}\hspace{0.7mm}v_{s_1s_2\cdots s_k}^{(t)}\hspace{0.5mm}v_{s'_1s'_2 \cdots s'_k}^{(t)}\vspace{2mm}\\
&
=\sum\limits_{i=0}^{n-1}{\sum\limits_{\substack{{s_1,s_2,\cdots,s_k }\\ {(s_1,s_2,\cdots,s_k)= (s'_1,s'_2,\cdots,s'_k)}}}}a_{(s_1s_2\cdots s_k)}^{(i)}\hspace{0.5mm}b_{(s_1s_2\cdots s_k)}^{(i)}
\Big(\sum\limits_{t=1}^{m_1m_2\cdots m_k}\hspace{0.5mm}(v_{(s_1s_2\cdots s_k)}^{(t)})^2\Big)\\&+ {\sum\limits_{i=0}^{n-1}}\hspace{0.5mm}{\sum\limits_{s_1,s_2,\cdots,s_k }}\hspace{-6mm}{\sum\limits_{\substack{{s'_1,s'_2,\cdots,s'_k }\\ {(s_1,s_2,\cdots,s_k)\neq (s'_1,s'_2,\cdots,s'_k)}}}}\hspace{-6mm}a_{s_1s_2\cdots s_k}^{(i)}\hspace{0.5mm}
b_{s'_1s'_2\cdots s'_k}^{(i)} \Big(\sum\limits_{t=1}^{m_1m_2\cdots m_k} v_{s_1s_2\cdots s_k}^{(t)} v_{s'_1s'_2\cdots s'_k}^{(t)}\Big )\vspace{3mm}\\&=\beta {\sum\limits_{i=0}^{n-1}}~\sum\limits_{s_1s_2\cdots s_k}a_{s_1s_2\cdots s_k}^{(i)}\hspace{0.5mm}b_{s_1s_2\cdots s_k}^{(i)}\vspace{2mm}\\&=\beta \sum\limits_{s_1s_2\cdots s_k}\Big(\sum\limits_{i=0}^{n-1}a_{s_1s_2\cdots s_k}^{(i)}\hspace{0.5mm}b_{s_1s_2\cdots s_k}^{(i)}\Big)=0,\end{array}$$
which proves the result. \hfill $\Box$

\section{Quantum codes from constacyclic codes over the ring $\mathcal{R}$}

Let $H$ be the Hilbert space of dimension $K$ over the field of complex numbers $\mathbb{C}$. Define $H^{\otimes n}$ to be the $n-$fold tensor product of the Hilbert Space $H$, i.e., $H^{\otimes n}=H\otimes H\otimes \cdots \otimes H$ ($n$ times). Then $H^{\otimes n}$ is a Hilbert Space of dimension $K^n$. A quantum code of length $n$ and dimension $K$ over $\mathbb{F}_q$ is defined to be the Hilbert subspace of $H^{\otimes n}$.
	
 A quantum code of length $n$, dimension $\ell$ and minimum distance $d$ is represented as $[[n,\ell,d]]_q$. Each quantum code satisfies the singleton bound, i.e., $\ell+2d \leq n+2$. If for some quantum code, singleton bound is attained, then it is said to be quantum MDS (Maximum Distance Seperable) code. It is said to be Almost MDS, if it satisfies $n=\ell+2d$. A quantum code $[[n,\ell,d]]_q$ is better than the other quantum code $[[n',\ell',d']]_q$, if any of the following condition holds:
	\begin{enumerate}[(a)]
		\item $\frac{\ell}{n} > \frac{\ell'}{n'}$, where $d=d'$ (greater code rate with same distance).
		\item $d>d'$, where $\frac{\ell}{n} = \frac{\ell'}{n'}$ (greater distance with same code rate).
	\end{enumerate}

\begin{lem}\label{CSS}\normalfont (\cite{Calderbank}) [CSS construction]
	If $\mathcal{C}$ is an $[n,\ell,d]$ linear code with $\mathcal{C}^\perp \subseteq \mathcal{C}$ over $\mathbb{F}_q$, then there exists a Quantum Error-Correcting code with parameters $[[n,2\ell-n,d]]$ over $\mathbb{F}_q$.
\end{lem}

\begin{theo}\label{th7}
For $q=p^e$ and any $n$  with $\gcd (n,p)\neq 1,$ there always exist Quantum MDS codes $[[n,n-2,2]]_q$.
\end{theo}

\noindent {\textbf{Proof:}}  Let $\mathcal{C}$ be the even weight linear code having parameters $[n,n-1,2]$ over $\mathbb{F}_q$ generated by $g(x)=x-1$. Here, $g^\perp(x)=-(x-1)$. $\mathcal{C}^\perp$ is generated by $h(x)=h^\perp (x)=1+x+x^2+\cdots +x^{n-1}$. As $p| n$, $1$ is a root of $h(x)$, so $x-1$ divides $h(x)$. Therefore, $$x^n-1 =g(x)h(x) \equiv 0\pmod {g(x)g^\perp(x)}.$$ By Theorem \ref{thm1}, $\mathcal{C}^\perp \subseteq \mathcal{C}$. Now, by CSS construction (Lemma \ref{CSS}), there exists Quantum MDS code $[[n,n-2,2]]_q$.

\begin{rk}\normalfont
For $\gcd (n,p) \neq 1$, the Quantum MDS code $[[n,n-2,2]]_q$ is better than many Quantum codes given in the database \cite{database}.\vspace{2mm}

 \noindent For example, the Quantum MDS code $[[60,58,2]]_5$ is better than $[[60,54,2]]_5$ as obtained in \cite{constajaa}, \cite{consta_22}; the Quantum MDS code $[[180,178,2]]_5$ is better than $[[180,156,2]]_5$ and $[[180,176,2]]_5$ obtained in \cite{AM} and \cite{DinIEEE} respectively; the Quantum MDS code $[[100,98,2]]_5$ is better than $[[100,92,2]]_5$ and $[[100,94,2]]_5$ obtained in \cite{ashraf2016} and \cite{BagIEEE} respectively; the Quantum MDS code $[[120,118,2]]_5$ is better than 
 $[[120,114,2]]_5$ obtained in
  \cite{constajaa}, \cite{consta_22} and \cite{MaGao};
 the Quantum MDS code $[[140,138,2]]_5$ is better than $[[140,112,2]]_5$ and $[[140,134,2]]_5$ obtained in \cite{ashraf2016} and \cite{BagIEEE} respectively.

\end{rk}

\begin{theo}\label{thm7}
		 Let $\mathcal{C}=\underset{s_1,s_2,\cdots,s_k}{\bigoplus}~\eta_{s_1s_2\cdots s_k}\mathcal{C}_{s_1s_2\cdots s_k}$ be a $\lambda-$ constacyclic code of length $n$ over $\mathcal{R}$. Suppose  $\Phi(\mathcal{C})$ has parameters $[m_1m_2\cdots m_k n,\ell,d_H]$. If $\mathcal{C}^\perp \subseteq \mathcal{C}$, then there exists a Quantum Error-Correcting code $[[m_1m_2\cdots m_k n,2\ell-m_1m_2\cdots m_k n,d_H]]_q$.
	\end{theo}
	
\noindent	Proof follows by using CSS construction {\cite{Calderbank}}, and Theorems \ref{thm1}, \ref{graydual}.

\subsection{Examples}

\noindent In this section, we obtain many optimal and better quantum codes than already existing  quantum codes. The MAGMA computation system is used to manage all of the computations in these examples. See  Tables 1, 2 and 3.

\begin{eg}\normalfont
	Let $\mathcal{R}=\mathbb{F}_5[u_1,u_2]/ \langle u_1^2-1,u_2^2-1,u_1u_2-u_2u_1 \rangle$  and $n=20$. \vspace{2mm}
	
\noindent Here $\eta_{11}= \frac{1}{4}(1+u_1+u_2+u_1u_2)    , \eta_{12}=\frac{1}{4}(1+u_1-u_2-u_1u_2)      ,\eta_{21}= \frac{1}{4}(1-u_1+u_2-u_1u_2)    $ and $\eta_{22}=\frac{1}{4}(1-u_1-u_2+u_1u_2) .$\vspace{2mm}

 \noindent Also $x^{20}-1=(x+1)^5(x+2)^5(x+3)^5(x+4)^5 \in \mathbb{F}_5[x]$. Take  $g_{11}(x)=(x+1)^2,~ g_{12}(x)=(x+2)^2,~ g_{21}(x)=x+3,~ g_{22}(x)=x+4$ and  $\mathcal{C}_{11}=\langle g_{11}(x)\rangle$, $\mathcal{C}_{12}=\langle g_{12}(x)\rangle$, $\mathcal{C}_{21}=\langle g_{21}(x)\rangle$, and $\mathcal{C}_{22}=\langle g_{22}(x)\rangle$.  Take
 $${\footnotesize V_1=\begin{bmatrix}
 	1&2&1&1\\
 	-2&1&1&-1\\
 	-1&-1&1&2\\
 	-1&1&-2&1
 	\end{bmatrix}}
 $$
 so that $V_1V_1^T=2I_4$.
\noindent  Then $\mathcal{C}=\eta_{11}\mathcal{C}_{11}\oplus \eta_{12}\mathcal{C}_{12}\oplus \eta_{21}\mathcal{C}_{21}\oplus \eta_{22}\mathcal{C}_{22}$  is a cyclic code of length $20$ and its Gray image $\Phi(\mathcal{C})$ has parameters $[80,74,3]$ over $\mathbb{F}_5$. We have $x^{20}-1\equiv 0 \pmod{g_{s_1s_2}(x)g^\perp_{s_1s_2}(x)}$ for $1\leq s_1,s_2\leq 2$. Hence by Theorem \ref{thm7}, there exists a quantum code $[[80,68,3]]_5$ which has the greater code rate than the code $[[80,56,3]]_5$ given by \cite{DinIEEE}.

\end{eg}

\begin{eg}\normalfont
	Let $\mathcal{R}=\mathbb{F}_{17}[u_1,u_2]/ \langle u_1^2-1,u_2^2-1,u_1u_2-u_2u_1 \rangle$  and $n=34$. \vspace{2mm}

	 \noindent Then $x^{34}+1=(x+4)^{17} (x+13)^{17} \in \mathbb{F}_{17}[x]$. Take  $g_{11}(x)=x^2+8x+16,~ g_{12}(x)=x+13,~ g_{21}(x)=1,~ g_{22}(x)=1$ and  $\mathcal{C}_{11}=\langle g_{11}(x)\rangle$, $\mathcal{C}_{12}=\langle g_{12}(x)\rangle$, $\mathcal{C}_{21}=\langle g_{21}(x)\rangle$, and $\mathcal{C}_{22}=\langle g_{22}(x)\rangle$.  Take
	 $${\footnotesize V_2=\begin{bmatrix}
	 	12&9&14&16\\
	 	9&5&16&3\\
	 	14&1&5&9\\
	 	16&14&8&5\\
	 	\end{bmatrix}}
	 $$
	 so that $V_2V_2^T=14I_4$.
	 \noindent  Then $\mathcal{C}=\eta_{11}\mathcal{C}_{11}\oplus \eta_{12}\mathcal{C}_{12}\oplus \eta_{21}\mathcal{C}_{21}\oplus \eta_{22}\mathcal{C}_{22}$  is a negacyclic code of length $34$ and its Gray image $\Phi(\mathcal{C})$ has parameters $[136,133,3]$ over $\mathbb{F}_{17}$. We have $x^{34}+1\equiv 0 \pmod{g_{s_1s_2}(x)g^\perp_{s_1s_2}(x)}$ for $1\leq s_1,s_2\leq 2$. Hence by Theorem \ref{thm7}, there exists a quantum code $[[136,130,3]]_{17}$ which is an Almost MDS and new quantum code as per the database \cite{database}.
	
\end{eg}

\begin{eg}\normalfont
	Let $\mathcal{R}=\mathbb{F}_{19}[u_1,u_2]/ \langle u_1^2-1,u_2,u_1u_2-u_2u_1 \rangle$  and $n=9$.
	 Here $\eta_{11}= \frac{1}{2}(1+u_1)$ and $\eta_{21}=\frac{1}{2}(1-u_1).$ \vspace{2mm}

	\noindent Then $x^{9}+1=(x+1)(x+4)(x+5)(x+6)(x+7)(x+9)(x+11)(x+16)(x+17)$ and $x^9-1=(x+2)(x+3)(x+8)(x+10)(x+12)(x+13)(x+14)(x+15)(x+18) \in \mathbb{F}_{19}[x]$. Take  $g_{11}(x)=x+7,~ g_{21}(x)=x+14$ and  $\mathcal{C}_{11}=\langle g_{11}(x)\rangle$, $\mathcal{C}_{21}=\langle g_{21}(x)\rangle$.  Take
	$${\footnotesize V_3=\begin{bmatrix}
		1&18\\
		1&1\\
		\end{bmatrix}}
	$$
	so that $V_3V_3^T=2I_2$.
	\noindent  Then $\mathcal{C}=\eta_{11}\mathcal{C}_{11}\oplus \eta_{21}\mathcal{C}_{21}$  is  $\eta_{21}-\eta_{11}=(-u_1)$- constacyclic code of length $9$ and its Gray image $\Phi(\mathcal{C})$ has parameters $[18,16,3]$ over $\mathbb{F}_{19}$. We have $x^{9}+1\equiv 0 \pmod{g_{s_1s_2}(x)g^\perp_{s_1s_2}(x)}$ for $s_1=s_2=1,$ and $x^{9}-1\equiv 0 \pmod{g_{s_1s_2}(x)g^\perp_{s_1s_2}(x)}$ for $s_1=2, s_2=1$. Hence by Theorem \ref{thm7}, there exists a quantum code $[[18,14,3]]_{19}$ which is an MDS  and  new  quantum code as per the database \cite{database}.
	
\end{eg}
\noindent The non-singular matrices $V$ taken for the construction the Quantum Codes are:

\[{\scriptsize V_4=\begin{bmatrix}
2&3&4\\
2&1&2\\
1&2&3\\
\end{bmatrix}_{\mathbb{F}_5}
~~~~~{ V_5=\begin{bmatrix}
1&28\\
1&1\\
\end{bmatrix}}_{\mathbb{F}_{29}}
~~~~~~V_6=\begin{bmatrix}
1&4\\
1&1\\
\end{bmatrix}_{\mathbb{F}_5}
~~~~~~V_7=\begin{bmatrix}
6&2\\
2&1\\
\end{bmatrix}_{\mathbb{F}_7}}\]

\[{\scriptsize
V_8=\begin{bmatrix}
1&6\\
1&1\\
\end{bmatrix}_{\mathbb{F}_7}
~~~~~~V_9=\begin{bmatrix}
10&2\\
2&1\\
\end{bmatrix}_{\mathbb{F}_{11}}
~~~~~~V_{10}=\begin{bmatrix}
3&3\\
3&10\\
\end{bmatrix}_{\mathbb{F}_{13}}
~~~~~V_{11}=\begin{bmatrix}
1&4\\
1&1\\
\end{bmatrix}_{\mathbb{F}_{25}}}\]

\[{\scriptsize
V_{12}=\begin{bmatrix}
1&2\\
1&1\\
\end{bmatrix}_{\mathbb{F}_{27}}
~~~~V_{13}=\begin{bmatrix}
1&6\\
1&1\\
\end{bmatrix}_{\mathbb{F}_{49}}
~~~~V_{14}=\begin{bmatrix}
1&10\\
1&1\\
\end{bmatrix}_{\mathbb{F}_{121}}
~~~~V_{15}=\begin{bmatrix}
1&16\\
1&1\\
\end{bmatrix}_{\mathbb{F}_{289}}
}\]

\[{\scriptsize
V_{16}=\begin{bmatrix}
9&2&1\\
10&9&2\\
2&1&2\\
\end{bmatrix}_{\mathbb{F}_{11}}
~~~~~V_{17}=\begin{bmatrix}
11&2&1\\
12&11&2\\
2&1&2\\
\end{bmatrix}_{\mathbb{F}_{13}}
~~~~~~~~~V_{18}=\begin{bmatrix}
2&1&2\\
15&2&1\\
1&2&15\\
\end{bmatrix}_{\mathbb{F}_{17}}}\]

\[{\scriptsize {V_{19}=\begin{bmatrix}
		10&7&9&8\\
		7&3&8&4\\
		9&5&3&7\\
		8&9&6&3\\
		\end{bmatrix}}_{\mathbb{F}_{13}}
~~~~~~~~~{V_{20}= \begin{bmatrix}
1&12\\
1&1
\end{bmatrix}}_{\mathbb{F}_{169}}}\]

\vspace{2mm}

\noindent In Table 1, we have constructed Quantum error-correcting codes from cyclic codes over $\mathcal{R}$, in Table 2, we have constructed Quantum error-correcting codes from negacyclic codes over $\mathcal{R}$.  In Table 3, we have constructed Quantum error-correcting codes from $(-\eta_{11}+\eta_{21})$-constacyclic codes over $\mathcal{R}$.  We have compared our codes with those obtained in some recent papers and this is  mentioned in `Remarks' column. Here, NQC refers to New Quantum Codes as per the database \cite{database}.
We express the generator polynomials  $g_{s_1\cdots s_k}(x)$ by their coefficients in the decreasing order, for example we denote a polynomial $x^4+10x^3+2x^2+0x+\zeta$ by $\zeta 02(10)1,$ where $\zeta$ denotes the primitive element of the field.

{\scriptsize   \begin{center} {\normalsize \textbf{Table 1}}\vspace{5mm}\\
		\begin{tabular}{|c|c|c|c|c|c|c|c|c|}
			\hline
			$n$ & $q$ &V& $f_1(u_1)$ & $f_2(u_2)$ & $g_{s_1\cdots s_k}(x)$ & $\Phi(\mathcal{C})$ &  $[[n,\ell,d]]_q$ &Remarks\\&&&&&&&&\\
			\hline
			
			8 & 5 &$V_4$& $u_1^3-u_1$ & $u_2$ & $21,1,1$ & $[24,23,2]$ & $[[24,22,2]]_5$ &NQC \\
			&&&&&&&MDS&\\
			\hline
			
			11&5 &$V_4$& $u_1^3-u_1$&$u_2-1$&$1,411421,$&$[33,23,3]$&$[[33,13,3]]_5$ &NQC \\&&&&&$431441$ &&&\\
			\hline
			
			20 & 5 &$V_1$& $u_1^2-1$ & $u_2^2-1$ & $121,441$ & $[80,74,3]$ & $[[80,68,3]]_5$ &$[[80,56,3]]_5$ \\
			&&&&&$31,41$&&&\cite{DinIEEE}\\
			\hline

			93 & 5 &$V_1$& $u_1^2-u_1$ & $u_2^2-1$ & $4311,$ & $[372,354,4]$ & $[[372,336,4]]_5$ &NQC \\
			&&&&&$1300341,$&&&\\&&&&&$4111,4201$&&&\\
			\hline

			105 & 5 &$V_6$& $u_1^2+u_1$ & $u_2+1$ & $1,111$  & $[210,208,2]$ & $[[210,206,2]]_{5}$ &$[[210,150,2]]_5$  \\
			&&&&&  &&Almost MDS&$[[210,204,2]]_5$\\&&&&&  &&&\cite{AM},\cite{constajaa}\\
			\hline
			
			108 & 5 &$V_6$& $u_1^2-1$ & $u_2$ & $31,1$  & $[216,215,2]$ & $[[216,214,2]]_{5}$ & $[[216,210,2]]_5$ \\
			&&&&&  &&MDS&\cite{constajaa}\\
			\hline
			
			63 & 7 &$V_8$& $u_1^2+u_1$ & $u_2-6$ & $325261,$  & $[126,120,3]$ & $[[126,114,3]]_{7}$  &$[[126,110,3]]_7$\\
			&&&&& $51$ &&&\cite{constajaa}\\
			\hline
			
			84 & 7 &$V_8$& $u_1^2-1$ & $u_2-3$ & $26361,$  & $[168,163,3]$ & $[[168,158,3]]_{7}$&$[[168,156,3]]_7$  \\
			&&&&& $31$ &&&\cite{constajaa}\\
			\hline
			
			84 & 7 &$V_7$& $u_1^2-u_1$ & $u_2+3$ & $661021,$ & $[168,160,4]$ & $[[168,152,4]]_7$ &NQC\\
			&&&&&$4411$&&&\\
			\hline
			
		11 & 11 &$V_9$& $u_1^2-1$ & $u_2-8$ & $(10)1,191$ & $[22,19,3]$ & $[[22,16,3]]_{11}$  &NQC\\
		&&&&&&&Almost MDS&\\
			\hline
			
			11 & 11 &$V_9$& $u_1^2-u_1$ & $u_2-3$ & $(10)1,(10)381$ & $[22,18,4]$ & $[[22,14,4]]_{11}$ & NQC\\
			&&&&&&&Almost MDS&\\
			\hline
			
			11 & 11 &$V_9$& $u_1$ & $u_2^2-1$ & $191,17671$ & $[22,16,5]$ & $[[22,10,5]]_{11}$ &NQC \\
			&&&&&&&&\\
			\hline
					
	\end{tabular}
	\end{center}}
	\noindent Contd(Table 1)...\newpage

			{\scriptsize   \begin{center}
					\begin{tabular}{|c|c|c|c|c|c|c|c|c|}
						\hline
			11 & 11 &$V_9$& $u_1-1$ & $u_2^2-u_2$ & $191,$ & $[22,15,6]$ & $[[22,8,6]]_{11}$ &$[[30,10,6]]_{11}$ \\
			&&&&&$(10)51(10)61$&&&\cite{consta35}\\
			\hline
			
			15 & 11 &$V_9$& $u_1^2+u_1$ & $u_2+1$ & $428791,$ & $[30,23,5]$ & $[[30,16,5]]_{11}$  &$[[30,10,5]]_{11}$\\
			&&&&&$87(10)1$&&&\cite{Koroglu}\\
			\hline

			33 & 11 &$V_9$& $u_1$ & $u_2^2+u_2$ & $(10)381,111$ & $[66,61,4]$ & $[[66,56,4]]_{11}$&$[[66,54,4]]_{11}$,  \\
			&&&&&&&&$[[66,52,4]]_{11}$\\&&&&&&&& \cite{consta35}~\cite{MaGao}\\
			\hline

			13 & 13 &$V_{10}$& $u_1$ & $u_2^2-1$ & $1(11)1,(12)1$ & $[26,23,3]$ & $[[26,20,3]]_{13}$& $[[26,18,3]]_{13}$~  \\
			&&&&&&&Almost MDS&\cite{consta35}\\
			\hline
			
	5 & 25 &$V_{11}$& $u_1^2-1$ & $u_2-\zeta$ & $131,41$ & $[10,7,3]$ & $[[10,4,3]]_{25}$  &NQC\\
			&&&&&&&Almost MDS&\\
			\hline
			
			7 & 25 &$V_{11}$& $u_1^2+1$ & $u_2-1$ & $4(\zeta^{13})(\zeta^5)1,$ & $[14,8,5]$ & $[[14,2,5]]_{25}$ &NQC \\
			&&&&&$4(\zeta^{17})(\zeta) 1$&&& \\
			\hline
			
			15 & 25 &$V_{11}$& $u_1^2-u_1$ & $u_2+1$ & $(\zeta^4)(\zeta^{15})(\zeta^7)1,$ & $[30,24,4]$ & $[[30,18,4]]_{25}$  &NQC\\
			&&&&&$4(\zeta^{10})(\zeta^{14})1$&&&\\
			\hline
				
	\end{tabular}
\end{center}}

{\scriptsize   \begin{center} {\normalsize \textbf{Table 2}}\vspace{5mm}\\
		\begin{tabular}{|c|c|c|c|c|c|c|c|c|}
			\hline
			$n$ & $q$ &V& $f_1(u_1)$ & $f_2(u_2)$ & $g_{s_1\cdots s_k}(x)$ & $\Phi(\mathcal{C})$ &  $[[n,\ell,d]]_q$&Remarks \\&&&&&&&&\\
			\hline
			
			77 & 11 &$V_{16}$& $u_1^3-u_1$ & $u_2$ & $121,$ $1,$ & $[231,226,3]$ & $[[231,221,3]]_{11}$&NQC\\
			&&&&&$1461$&&&\\
			\hline
			
			99 & 11 &$V_{16}$& $u_1^3-u_1$ & $u_2+3$ & $121,$ $1,$  & $[297,289,3]$ & $[[297,281,3]]_{11}$ &NQC \\
			&&&&& $100(10)001$ &&&\\
			\hline

			126 & 11 &$V_{16}$& $u_1^3-u_1$ & $u_2-1$ & $14(10)1701, $  & $[378,366,3]$ & $[[378,354,3]]_{11}$&NQC  \\
			&&&&& \hspace{-3mm}1,158(10)471\hspace{-3mm} &&&\\
			\hline
			
			154 & 11 &$V_{16}$& $u_1^3-u_1$ & $u_2-5$ & \hspace{-3mm}10806040601,\hspace{-3mm}  & $[462,450,3]$ & $[[462,438,3]]_{11}$ &NQC \\
			&&&&&$101,$ $1$ &&&\\
			\hline
						
			52 & 13 &$V_{17}$& $u_1^3-u_1$ & $u_2$ & $(12)0(10)01, $ & $[156,150,3]$ & $[[156,144,3]]_{13}$ &NQC \\
			&&&&&801, 1&&&\\
			\hline
			52 & 13 &$V_{19}$& $u_1^2-u_1$ & $u_2^2-2u_2$ & $1, 1, 1, 801$ & $[208,206,2]$ & $[[208,204,2]]_{13}$ &NQC\\
			&&&&&&&Almost MDS&\\
			\hline

			78 & 13 &$V_{17}$& $u_1^3-u_1$ & $u_2-9$ & $(12)(10)1, $ & $[234,230,3]$ & $[[234,226,3]]_{13}$ &NQC \\
			&&&&&61, 71&&&\\
			\hline
			
			78 & 13 &$V_{19}$& $u_1^2-u_1$ & $u_2^2-u_2$ & $(12)(10)1,$ & $[312,308,3]$ & $[[312,304,3]]_{13}$ &$[[312,264,3]]_{13}$
			\\&&&&&61,71, 1&&& $[[312,288,3]]_{13}$\\&&&&&&&&\cite{consta29},\cite{DinIEEE}\\	
			\hline
			
			91 & 13 &$V_{17}$& $u_1^3-u_1$ & $u_2-4$ & $121,$ $1,$ $181$ & $[273,269,3]$ & $[[273,265,3]]_{13}$&NQC \\
			&&&&&&&&\\
			\hline			
			
			91 & 13 &$V_{19}$& $u_1^2-u_1$ & $u_2^2+u_2$ & $121, 1, 171, $ & $[364,359,3]$ & $[[364,354,3]]_{13}$ &NQC \\
			&&&&&11&&&\\
			\hline
			
			34 & 17 &$V_{2}$& $u_1^2-1$ & $u_2^2-1$ & $(16)81, $ & $[136,133,3]$ & $[[136,130,3]]_{17}$ &NQC \\
			&&&&&(13)1, 1, 1&&Almost MDS&\\
			\hline
			
			51 & 17 &$V_{18}$& $u_1^3-u_1$ & $u_2$ & $121,$ & $[153,149,3]$ & $[[153,145,3]]_{17}$&NQC\\&&&&&$1(16)1, 1$&&& \\
			\hline	
			
			51 & 17 &$V_{2}$& $u_1^2+u_1$ & $u_2^2-1$ & $(16)81, 1, 1,$ & $[204,200,3]$ & $[[204,196,3]]_{17}$& NQC \\
			&&&&&$1(16)1$&&&\\
			\hline			
			
			18 & 27 &$V_{12}$& $u_1^2-1$ & $u_2+1$ & $101, 1$  & $[36,34,2]$ & $[[36,32,2]]_{27}$ &$[[36,28,2]]_{27}$ \\
			&&&&&&&Almost MDS&\cite{consta_8}\\
			\hline
						
			28 & 49 &$V_{13}$& $u_1^2-1$ & $u_2-5$ & $2121,(\zeta^{42})1$ & $[56,52,3]$ & $[[56,48,3]]_{49}$ &$[[56,40,3]]_{49}$ \\
			&&&&&&&&\cite{consta_8}\\
			\hline
			
			44 & 121 &$V_{14}$& $u_1^2-1$ & $u_2-9$ & $(10)81,\zeta^{45}1$ & $[88,84,3]$ & $[[88,80,3]]_{121}$ &$[[88,72,3]]_{121}$ \\
			&&&&&&&&\cite{consta_8}\\
			\hline
			
			18 & 169 &$V_{20}$& $u_1^2-1$ & $u_2+1$ & $2001,9(12)1$ & $[36,31,3]$ & $[[36,26,3]]_{169}$&NQC\\
			&&&&&&&&\\
			\hline

			18 & 169 &$V_{20}$& $u_1^2-1$ & $u_2+1$ & $21,1$ & $[36,35,2]$ & $[[36,34,2]]_{169}$&NQC\\
			&&&&&&&MDS&\\
			\hline
			
			52 & 169 &$V_{20}$& $u_1^2-1$ & $u_2-11$ & $(\zeta^{21})6(\zeta^{77})1,$ & $[104,100,3]$ & $[[104,96,3]]_{169}$&$[[104,88,3]]_{169}$  \\
			&&&&&$(\zeta^{63})1$&&&\cite{consta_8}\\
			\hline
						
			12 & 289 &$V_{15}$& $u_1^2-1$ & $u_2$ & $(\zeta^{12})1,$ $1$  & $[24,23,2]$ & $[[24,22,2]_{289}$& NQC \\
			&&&&&&& MDS&\\
			\hline
					
		\end{tabular}
	\end{center}}\vspace{3mm}
\newpage
	{\scriptsize   \begin{center} {\normalsize \textbf{Table 3}}\vspace{5mm}\\
			\begin{tabular}{|c|c|c|c|c|c|c|c|c|}
				\hline
				$n$ & $q$ &V& $f_1(u_1)$ & $f_2(u_2)$ & $g_{s_1\cdots s_k}(x)$ & $\Phi(\mathcal{C})$ &  $[[n,\ell,d]]_q$&Remarks \\&&&&&&&&\\
				\hline
				
				7 & 7 &$V_{8}$& $u_1^2-1$ & $u_2+1$ & $121, 61$ & $[14,11,3]$ & $[[14,8,3]]_{7}$ &NQC \\
				&&&&&&&Almost MDS&\\
				\hline
				
				14 & 7 &$V_{8}$& $u_1^2-u_1$ & $u_2-2$ & $101, 151$ & $[28,24,3]$ & $[[28,20,3]]_{7}$ &NQC \\
				&&&&&&&&\\
				\hline
				
				6 & 19 &$V_{3}$& $u_1^2+u_1$ & $u_2-17$ & $701, 71$ & $[12,9,3]$ & $[[12,6,3]]_{19}$ &NQC \\
				&&&&&&&Almost MDS&\\
				\hline
				
				9 & 19 &$V_{3}$& $u_1^2-1$ & $u_2$ & $71, (14)1$ & $[18,16,3]$ & $[[18,14,3]]_{19}$ &NQC \\
				&&&&&&&MDS&\\
				\hline
				
				4 & 29 &$V_{5}$& $u_1^2-u_1$ & $u_2$ & $(12)01, (12)1$ & $[8,5,3]$ & $[[8,2,3]]_{29}$ &NQC \\
				&&&&&&&Almost MDS&\\
				\hline
				
				12 & 29 &$V_{5}$& $u_1^2-1$ & $u_2-13$ & $(12)01, (28)(12)1$ & $[24,21,3]$ & $[[24,18,3]]_{29}$ &NQC \\
				&&&&&&&Almost MDS&\\
				\hline
				
				14 & 29 &$V_{5}$& $u_1^2+u_1$ & $u_2+1$ & $(24)(11)1, 71$ & $[28,25,3]$ & $[[28,22,3]]_{29}$ &NQC \\
				&&&&&&&Almost MDS&\\
				\hline

			\end{tabular}
		\end{center}}

\section{Irrelevance of the polynomials $f_i(u_i)$}\label{sec5}
	
\noindent While constructing quantum codes, we observe that  the choice of the polynomials $f_i(u_i)$ is irrelevant, it depends only on their degrees $m_i$ and on the non-singular matrix $V$ taken in the definition of Gray map $\Phi$, as is seen in the following theorem and corollary.	 
	
		\begin{theo}\label{thm8} Let for each $i, 1\leq i \leq k $, $f_i(u_i)$ and $f'_i(u_i)$ be two sets of polynomials of the same degree $m_i$ which split into distinct linear factors over $\mathbb{F}_q$. Let $\mathcal{R}=\mathbb{F}_{q}[u_1,u_2, \cdots, u_k]/\langle f_i(u_i),u_iu_j-u_ju_i\rangle$ and  $\mathcal{R}'=\mathbb{F}_{q}[u_1,u_2, \cdots, u_k]/\langle f'_i(u_i),\\u_iu_j-u_ju_i\rangle$ be two different non-chain rings.  Let $\eta_{s_1s_2\cdots s_k},~ \eta'_{s_1s_2\cdots s_k}$ for $1\leq s_i\leq m_i, 1 \leq i \leq k$ be the corresponding primitive central idempotents of the rings $\mathcal{R}$ and $\mathcal{R'}$. Suppose

 \begin{enumerate}[$\rm(i)$]\item  $\mathcal{C}_{s_1s_2\cdots s_k}$ are some $\lambda_{s_1s_2\cdots s_k}$   constacyclic codes over $\mathbb{F}_q$,
 	 \item  $~\mathcal{C}=\underset{s_1,s_2,\cdots,s_k}{\bigoplus}~\eta_{s_1s_2\cdots s_k}\mathcal{C}_{s_1s_2\cdots s_k}$ is a $\lambda = \underset{s_1,s_2,\cdots,s_k}{\bigoplus}~ \eta_{s_1s_2\cdots s_k}\lambda_{s_1s_2\cdots s_k}$ constacyclic code of length $n$ over $\mathcal{R}$,
 	   \item $\mathcal{C}'=\underset{s_1,s_2,\cdots,s_k}{\bigoplus}~\eta'_{s_1s_2\cdots s_k}\mathcal{C}_{s_1s_2\cdots s_k}$ is a $\lambda'=\underset{s_1,s_2,\cdots,s_k}{\bigoplus}\eta'_{s_1s_2\cdots s_k}\lambda_{s_1s_2\cdots s_k}$ constacyclic code of length $n$ over  $\mathcal{R}'$ and
 	   \item  Further let $\Phi$ and $\Phi'$ be the corresponding Gray maps on $\mathcal{R}$ and $\mathcal{R'}$; the non-singular matrix $V$ being the same for both $\Phi$ and $\Phi'$. \end{enumerate}

 \noindent  Then the codes $$\Phi(\mathcal{C})=\Phi'(\mathcal{C'}).$$
					
		\end{theo}
		
		\noindent \textbf{Proof:} Let $G_{s_1s_2\cdots s_k}$ be the generator matrices of the codes $\mathcal{C}_{s_1s_2\cdots s_k}$ for $1\leq s_i\leq m_i, 1 \leq i \leq k$. Then generator matrices of $\mathcal{C}$ and $\mathcal{C'}$ are \vspace{2mm}

		\[G=\begin{bmatrix}
		\eta_{11\cdots 1}G_{11\cdots 1} \\
		\eta_{21\cdots 1}G_{21\cdots 1}  \\
		\cdots \\
		\eta_{m_11\cdots 1}G_{m_11\cdots 1} \\
		\eta_{1 2 1\cdots 1}G_{12 1\cdots 1} \\
		\cdots \\
		\eta_{m_12 1\cdots 1}G_{m_12 1\cdots 1} \\
		\cdots \\
		\eta_{1m_2 1\cdots 1}G_{1m_2 1\cdots 1} \\\cdots\\
		\cdots\\\eta_{1m_2 \cdots m_k}G_{1m_2 \cdots m_k} \\\cdots\\
		\eta_{m_1m_2\cdots m_k}G_{m_1m_2\cdots m_k} \\
		\end{bmatrix}
		~~~~~~~~~G'=\begin{bmatrix}
		\eta'_{11\cdots 1}G_{11\cdots 1} \\
		\eta'_{21\cdots 1}G_{21\cdots 1}  \\
		\cdots \\
		\eta'_{m_11\cdots 1}G_{m_11\cdots 1} \\
		\eta'_{1 2 1\cdots 1}G_{12 1\cdots 1} \\
		\cdots \\
		\eta'_{m_12 1\cdots 1}G_{m_12 1\cdots 1} \\
		\cdots \\
		\eta'_{1m_2 1\cdots 1}G_{1m_2 1\cdots 1} \\\cdots\\
		\cdots\\\eta'_{1m_2 \cdots m_k}G_{1m_2 \cdots m_k} \\\cdots\\
		\eta'_{m_1m_2\cdots m_k}G_{m_1m_2\cdots m_k} \\
		\end{bmatrix}\]\vspace{2mm}
		
		\noindent Since, for $a \in \mathbb{F}_q$, $\Phi(a \eta_{s_1s_2\cdots s_k})=a \Phi(\eta_{s_1s_2\cdots s_k})=a.(s_1s_2\cdots s_k^{th} \mbox{ row of V})$ and, $\Phi'(a \eta'_{s_1s_2\cdots s_k})=a \Phi'(\eta'_{s_1s_2\cdots s_k})=a.(s_1s_2\cdots s_k^{th} \mbox{~row of V})$. Therefore, $\Phi(a \eta_{s_1s_2\cdots s_k})=\Phi'(a \eta'_{s_1s_2\cdots s_k})$. And hence,
		
		\[\Phi(G)=\begin{bmatrix}
		\Phi(\eta_{11\cdots 1}G_{11\cdots 1}) \\
		\Phi(\eta_{21\cdots 1}G_{21\cdots 1})  \\
		\cdots \\
		\Phi(\eta_{m_11\cdots 1}G_{m_11\cdots 1}) \\
		\Phi(\eta_{1 2 1\cdots 1}G_{12 1\cdots 1}) \\
		\cdots \\
		\Phi(\eta_{m_12 1\cdots 1}G_{m_12 1\cdots 1}) \\
		\cdots \\
		\Phi(\eta_{1m_2 1\cdots 1}G_{1m_2 1\cdots 1}) \\\cdots\\
		\cdots\\
		\Phi(\eta_{1m_2 \cdots m_k}G_{1m_2 \cdots m_k}) \\ \cdots\\
		\Phi(\eta_{m_1m_2\cdots m_k}G_{m_1m_2\cdots m_k}) \\
		\end{bmatrix}
		=\Phi'(G')=\begin{bmatrix}
		\Phi'(\eta'_{11\cdots 1}G_{11\cdots 1}) \\
		\Phi'(\eta'_{21\cdots 1}G_{21\cdots 1})  \\
		\cdots \\
		\Phi'(\eta'_{m_11\cdots 1}G_{m_11\cdots 1}) \\
		\Phi'(\eta'_{1 2 1\cdots 1}G_{12 1\cdots 1}) \\
		\cdots \\
		\Phi'(\eta'_{m_12 1\cdots 1}G_{m_12 1\cdots 1}) \\
		\cdots \\
		\Phi'(\eta'_{1m_2 1\cdots 1}G_{1m_2 1\cdots 1}) \\\cdots\\
		\cdots\\
		\Phi'(\eta'_{1m_2 \cdots m_k}G_{1m_2 \cdots m_k}) \\ \cdots\\
		\Phi'(\eta'_{m_1m_2\cdots m_k}G_{m_1m_2\cdots m_k}) \\
		\end{bmatrix}\]

		\noindent Therefore, the codes $\Phi(G)$ and $\Phi'(G')$ give rise to same codes over $\mathbb{F}_q$.
		
\begin{cor} \normalfont Take $\lambda_{s_1s_2\cdots s_k}=\pm 1$, $\lambda = \underset{s_1,s_2,\cdots,s_k}{\bigoplus}~\pm \eta_{s_1s_2\cdots s_k}$ and $\lambda' = \underset{s_1,s_2,\cdots,s_k}{\bigoplus}~\pm \eta'_{s_1s_2\cdots s_k}$ in the above Theorem \ref{thm8}. Suppose further $\mathcal{C}_{s_1s_2\cdots s_k}^\perp \subseteq \mathcal{C}_{s_1s_2\cdots s_k}$ for all  $s_i, ~1\leq s_i\leq m_i, 1 \leq i \leq k$, so that $\mathcal{C}^\perp \subseteq \mathcal{C}$ and $\mathcal{C}'^\perp \subseteq \mathcal{C}'$. Since by Theorem \ref{thm8}, $\Phi(\mathcal{C})= \Phi'(\mathcal{C}')$, we find that  $\Phi(\mathcal{C})$ and $ \Phi'(\mathcal{C}')$ give the same quantum-error correcting code using  Theorem \ref{thm7}.  \end{cor}

	\section{Conclusion}
	We study quantum codes over a finite field $\mathbb{F}_q$ from constacyclic codes over a  finite non-chain ring $\mathcal{R}=\mathbb{F}_{q}[u_1,u_2, \cdots, u_k]/\langle f_i(u_i),u_iu_j-u_ju_i\rangle$, where $f_i(u_i)$  are polynomials,  which split into distinct linear factors over $\mathbb{F}_{q}$. As a consequence, some new and better quantum codes as compared to the best known codes are obtained. We have also proved that, if we start with codes over the field $\mathbb{F}_q$, then using CSS construction, the construction of Quantum codes from constacyclic codes over $\mathcal{R}$ is independent of the choice of the polynomials $f_i(u_i)$, it depends only on their degrees. We also show that there always exists Quantum MDS code $[[n,n-2,2]]_q$ for any $n$  with $\gcd (n,q)\neq 1.$\\
	
\noindent \textbf{Acknowledgements:} The research of first author is funded by CSIR-UGC grant number 1074/(CSIR-UGC NET JUNE 2019). The third author is grateful to National Academy of Sciences, India for the financial support under Senior Scientist Platinum Jubilee Fellowship.

\end{document}